\documentclass[aps,pre,reprint,twocolumn,groupedaddress]{revtex4-2}

\usepackage{graphicx}      
\usepackage{amsmath}       
\usepackage{amsfonts}      
\usepackage{amssymb}       
\usepackage{bm}            
\usepackage{microtype}     
\usepackage{xcolor}
\usepackage{tikz}
\usetikzlibrary{arrows.meta,calc,positioning,backgrounds,shapes.geometric}

\definecolor{compBlue}{HTML}{1F77B4} 
\definecolor{compGreen}{HTML}{2CA02C} 
\definecolor{compOrange}{HTML}{FF7F0E} 
\definecolor{compRed}{HTML}{D62728} 
\definecolor{compPurple}{HTML}{9467BD} 
\definecolor{bgGray}{HTML}{F8F9FA} 

\definecolor{cKOne}{HTML}{3B73B9}
\definecolor{cKTwo}{HTML}{55A868}	
\definecolor{cKThree}{HTML}{D9842B}
\definecolor{cKOneTwo}{HTML}{C62828}
\definecolor{cKOneThree}{HTML}{7E57C2}
\definecolor{cKTwoThree}{HTML}{B2A429}
\definecolor{cKAll}{HTML}{2A9DAD}
\definecolor{cHeatHigh}{HTML}{BE1E1E}
\definecolor{cHeatMid}{HTML}{D08A8A}
\definecolor{cHeatLow}{HTML}{6E95C4}
		
\begin{document}
	
	\title{Moment-Resolved Readout and Reservoir Diversity in Nonequilibrium Langevin Computing}
	
	\author{Jizheng Duan}
	\email{duanjz@impcas.ac.cn}
	\author{MingYang Zhao}
	\author{YanWei Chen}
	\author{Lei Yang}
	\email{lyang@impcas.ac.cn} 
	\affiliation{Institute of Modern Physics, Chinese Academy of Sciences}
	\date{\today}

	\begin{abstract}
		Nonlinear thermodynamic computers based on Langevin dynamics exploit thermal fluctuations as a physical substrate for computation. Recent work has shown that quartic-confined fluctuating degrees of freedom can act as thermodynamic neurons capable of nonlinear function approximation at finite observation times. Here we extend this paradigm from mean-only readout to moment-resolved readout. Instead of representing each driven reservoir solely by its first moment, we construct a response vector from the elementwise raw polynomial moments
		\(\mathbb{E}[\bm{x}]\),
		\(\mathbb{E}[\bm{x}^{\odot 2}]\), and
		\(\mathbb{E}[\bm{x}^{\odot 4}]\).
		These observables combine displacement and central-shape contributions and are naturally aligned with the linear, quadratic, and quartic terms of the local driven dynamics.
		
		We further introduce a heterogeneous multi-reservoir architecture in which three reservoirs with distinct initialization and training histories form a joint \(2304\)-dimensional response representation. Under the fixed MNIST \(60000/10000\) reproduction protocol, feature-level fusion achieves the best observed accuracy of \(9695/10000=96.95\%\), compared with \(9682/10000=96.82\%\) for the strongest single-reservoir model and \(9684/10000=96.84\%\) for equal-weight logit averaging. An exact paired McNemar test does not establish a statistically significant improvement over the strongest single reservoir, but the ablation and wrong-set overlap results provide suggestive evidence of complementary classification errors. These results motivate higher-order polynomial-moment readout and reservoir heterogeneity as candidate design principles for finite-time Langevin computing.
	\end{abstract}

	\maketitle

	\section{Introduction}

	Thermodynamic computing has recently re-emerged as a promising physical route for information processing beyond conventional deterministic digital architectures~\cite{726791,proceedings2020047023,TANAKA2019100,wright_deep_2022,aifer_thermodynamic_2024,melanson_thermodynamic_2025,PhysRevX.11.021045,5392446,PhysRevE.110.054136,nrzn-h5ph,9450161,3045118.3045358}. Thermodynamic computers do not regard thermal fluctuations as sources of error to be suppressed, but rather seek to transform the random motion itself into a computational resource. Along this direction, Whitelam and Casert introduced a nonlinear thermodynamic computing framework: within this framework, fluctuating degrees of freedom constrained by a quartic potential and coupled to a thermal bath act as thermodynamic neurons capable of performing nonlinear computations~\cite{whitelam_nonlinear_2026}.
	
	In the Whitelam--Casert paradigm, the elementary computing unit is a continuously fluctuating physical coordinate whose activity is shaped by a nonlinear energy landscape. A driven quartic potential converts an external input into an input-dependent stochastic response, so that the observed state of the system at a finite time \(t_f\) implements an effective nonlinear activation. This idea provides a physically grounded alternative to abstract artificial neurons: the activation function is not prescribed analytically, but emerges from the nonequilibrium evolution of a thermally driven dynamical system. The resulting architecture establishes a direct connection between stochastic thermodynamics, nonlinear dynamical systems, and neural computation.
	
	However, the representational capacity of such a Langevin computer depends critically on how the fluctuating state is read out. If the final-time distribution of a thermodynamic neuron is reduced only to its mean displacement, the readout corresponds to a first-order coarse graining of the conditional probability density \(P(x,t_f|I)\). This is sufficient to define a deterministic activation-like response, but it discards much of the physical information carried by the nonequilibrium ensemble. In a nonlinear potential, the input does not merely shift the center of the distribution; it can also reshape its width, tail weight, asymmetry, and non-Gaussian structure. These changes are especially important for a quartic landscape, where the fourth-order confinement term directly controls the high-amplitude excursions of the stochastic coordinate.
	
	This observation motivates the central question of the present work: can a nonequilibrium thermodynamic computer be made more
	expressive by reading out not only the mean response, but an
	ordered set of distributional response statistics? We answer this question by introducing moment-resolved readout, a moment-based extension of Langevin computing. For each driven thermodynamic reservoir, instead of using only the first moment \(\mathbb{E}[\bm{x}]\), we construct an energy-aligned polynomial response vector
	\[
	\phi_{\bm K}(\bm I)
	=
	\left[
	\mathbb{E}[\bm{x}],
	\mathbb{E}[\bm{x}^{\odot 2}],
	\mathbb{E}[\bm{x}^{\odot 4}]
	\right],
	\]
	where \(\bm{x}^{\odot p}\) denotes the elementwise \(p\)-th power. The first moment captures the mean displacement response. The second and fourth raw moments provide progressively higher-order polynomial summaries of the finite-time distribution and are naturally aligned with the quadratic and quartic terms of the local confinement. Because they are raw rather than central moments, however, they mix changes in the distribution center with changes in its width, asymmetry, and higher-order shape. We therefore interpret them as energy-aligned polynomial response observables, rather than as pure measures of variance or non-Gaussianity.

	Recent progress has also shown that Langevin thermodynamic computers can be used for generative modeling, where structured data are synthesized from noise by the natural time evolution of a trained thermodynamic system~\cite{kwyy-1xln}. In that setting, the central quantity is the probability of generating the reverse of a noising trajectory, and the learning rule admits a physical interpretation in terms of heat emission and entropy production. This generative perspective demonstrates that nonequilibrium Langevin dynamics can encode data-producing transformations in an energy landscape. The present work addresses a complementary discriminative problem. Rather than asking how a thermodynamic computer can generate structured samples from noise, we ask how a finite-time nonequilibrium distribution should be read when the task is pattern recognition.
	
	The second limitation addressed in this paper concerns the use of a single physical reservoir. A single Langevin network can generate rich nonlinear responses, but the geometry of its responses is constrained by fixed input projections, coupling topology, training pathways, and noise scale~\cite{sekimoto_langevin_1998}. Simply increasing the size of a homogeneous reservoir does not guarantee a corresponding increase in useful computational diversity, as additional coordinates may explore regions of the response space that are highly correlated \cite{PhysRevE.60.2721,PhysRevE.92.012131}. For physical computing systems, this can lead to redundancy: two reservoirs may perform excellently as independent classifiers, but if their error patterns are too similar, combining them will not improve generalization ability.

	To overcome this limitation, we introduce a heterogeneous multi-reservoir architecture~\cite{WOO2024129334,Nakajima_2020,TANAKA2019100,Jaeger2001TheechoST,Maass2002RealTimeCW}. Instead of using identical replicas of one trained thermodynamic
	computer, we construct three reservoirs with different dynamical
	origins: a reservoir \(K_1\) trained directly using the
	polynomial-response surrogate, a second reservoir \(K_2\)
	initialized from a mean-response-trained configuration and
	subsequently refined using the polynomial surrogate, and a third
	reservoir \(K_3\) initialized independently with a stronger
	internal coupling scale~\cite{appeltant_information_2011,LUKOSEVICIUS2009127}. These reservoirs are designed to generate partially distinct response bases under the same input stimulus. The final representation is the feature-level concatenation
	\[
	\Phi(\bm{I})
	=
	\left[
	\phi_{\bm{K}_1}(\bm{I}),
	\phi_{\bm{K}_2}(\bm{I}),
	\phi_{\bm{K}_3}(\bm{I})
	\right],
	\]
	followed by a single standardized linear readout. The purpose of this construction is to examine whether reservoirs with distinct dynamical and training histories can provide complementary response coordinates and classification errors. Because the multi-reservoir model also increases feature dimension and computational resources, the present comparison should not by itself be interpreted as a complete separation of heterogeneity effects from capacity or sampling-budget effects. This work therefore extends nonlinear thermodynamic computing out of equilibrium in two complementary directions. First, it replaces mean-only readout with moment-resolved readout. Second, it replaces a single-reservoir representation with a heterogeneous multi-reservoir response basis, allowing weaker reservoirs with partially complementary error patterns to contribute to the joint representation~\cite{648054.743935,Hansen1990NeuralNE,annurev:/content/journals/10.1146/annurev-conmatphys-031119-050745}.

	\section{Methodology}
	
	\subsection{Stochastic Thermodynamics of Quartic-Confined Langevin Neurons}
	
	We model the thermodynamic computer as a collection of \(N\) coupled physical nodes driven by nonequilibrium stochastic dynamics. Given an external pattern vector \(\bm{I}\in\mathbb{R}^{D}\), the microstate of the \(r\)-th reservoir, \(r\in\{1,\dots,R\}\), is described by the time-dependent vector
	\(\bm{x}(t)=[x_1(t),x_2(t),\dots,x_N(t)]^\top\).
	Each nonlinear degree of freedom evolves according to the overdamped Langevin equation~\cite{PhysRevE.66.016119,gardiner2009stochastic}
	\begin{equation}
		\label{eq:langevin}
		dx_i=\mu F_i(\bm{x},\bm{I})\,dt+\sqrt{2\mu k_B T}\,dB_i(t),
	\end{equation}
	where \(\mu\) is the mobility parameter setting the intrinsic relaxation time scale, \(k_B T\) denotes the thermal noise intensity of the background heat bath, and \(B_i(t)\) are mutually independent standard Brownian motions. The deterministic force acting on the \(i\)-th node combines the local quartic confinement, internal coupling, static bias, and external input projection:
	\begin{equation}
		\label{eq:force}
		F_i(\bm{x},\bm{I})
		=
		-2J_2x_i-4J_4x_i^3
		+b_i+\sum_{j=1}^{N}W_{ij}x_j
		+\sum_{k=1}^{D}K_{ik}I_k .
	\end{equation}
	Here \(J_2\) and \(J_4\) specify the quadratic and stabilizing quartic components of the intrinsic potential, respectively. The matrices \(\bm{W}\) and \(\bm{K}\) define the internal recurrent interaction topology and the external input projection, while \(b_i\) is a local static bias. For an isolated node, or when discussing only the onsite contribution to the coupled dynamics, the input and bias act through the effective scalar drive
	\[
	h_i(\bm{I})=b_i+\sum_{k=1}^{D}K_{ik}I_k,
	\]
	leading to the local quartic potential
	\begin{equation}
		\label{eq:local_potential}
		U_i(x_i;h_i)=J_2x_i^2+J_4x_i^4-h_i x_i .
	\end{equation}
	
	In the numerical implementation, the recurrent matrix is directed and is not constrained to be symmetric. At initialization, its off-diagonal elements are independently sampled according to
	\begin{equation}
		W_{ij}\sim
		\mathcal{N}\!\left(0,\frac{s_W^2}{N}\right),
		\qquad i\neq j,
		\qquad
		W_{ii}=0 .
		\label{eq:w_initialization}
	\end{equation}
	The input and bias parameters are initialized as
	\begin{equation}
		K_{ik}\sim
		\mathcal{N}\!\left(0,\frac{s_K^2}{D}\right),
		\qquad
		b_i\sim\mathcal{N}(0,s_b^2),
		\label{eq:kb_initialization}
	\end{equation}
	with the values of \(s_W\), \(s_K\), and \(s_b\) specified in the numerical protocol below.
	
	A global scalar potential for the coupled system would require the integrability condition
	\begin{equation}
		\frac{\partial F_i}{\partial x_j}
		=
		\frac{\partial F_j}{\partial x_i},
		\qquad\text{equivalently}\qquad
		W_{ij}=W_{ji}.
		\label{eq:integrability_condition}
	\end{equation}
	If \(\bm W=\bm W^{\mathsf T}\), the deterministic force can be derived from
	\begin{equation}
		U(\bm{x};\bm I)
		=
		\sum_{i=1}^{N}
		\left(
		J_2x_i^2+J_4x_i^4-h_i(\bm I)x_i
		\right)
		-
		\frac{1}{2}\bm{x}^{\mathsf T}\bm W\bm{x}.
		\label{eq:global_symmetric_potential}
	\end{equation}
	The matrices used here are generally asymmetric, and therefore the full coupled dynamics are not assumed to derive from a global scalar energy landscape. In this paper, the term ``energy-sensitive'' refers specifically to local polynomial observables aligned with the onsite quadratic and quartic confinement, while the recurrent interaction acts as a generally nonconservative nonequilibrium drive.
	
	At the ensemble level, the finite-time probability density \(P(\bm{x},t|\bm{I})\) obeys the associated Fokker--Planck equation~\cite{Risken1996}
	\begin{equation}
		\label{eq:fokker_planck}
		\frac{\partial P}{\partial t}
		=
		-\mu\sum_{i=1}^{N}
		\frac{\partial}{\partial x_i}
		\left[
		F_i(\bm{x},\bm{I})P
		\right]
		+
		\mu k_B T
		\sum_{i=1}^{N}
		\frac{\partial^2 P}{\partial x_i^2}.
	\end{equation}
	Because the physical network is observed at a finite readout time \(t_f\), we do not assume that \(P(\bm{x},t_f|\bm{I})\) has relaxed to an equilibrium Boltzmann distribution.
	
	Rather than representing this finite-time nonequilibrium distribution only by the first moment
	\(\mathbb{E}[\bm{x}]\), we construct a finite-dimensional moment-resolved response representation from raw moment observables. For the coupled nonlinear Langevin system considered here, an exact analytical expression for the transient density \(P(\bm{x},t_f|\bm I)\) is generally unavailable. We therefore approximate its response statistics by Monte Carlo sampling of the stochastic dynamics. The theoretical \(p\)-th raw moment of the \(i\)-th coordinate is
	\begin{equation}
		\label{eq:theoretical_moment}
		\mathbb{E}\!\left[x_i^p(t_f)\mid \bm I\right]
		=
		\int x_i^p P(\bm{x},t_f|\bm I)\,d\bm{x}.
	\end{equation}
	Using \(M\) independently reset trajectories initialized from
	\(\bm{x}(0)=\bm 0\), this expectation is estimated by the empirical ensemble average
	\begin{equation}
		\label{eq:empirical_moment}
		m_{p,i}(\bm{I})
		=
		\frac{1}{M}\sum_{m=1}^{M}
		\left[
		x_i^{(m)}(t_f;\bm{I})
		\right]^p,
		\qquad
		p\in\{1,2,4\}.
	\end{equation}
	This estimator is the standard Monte Carlo approximation to the corresponding finite-time moment of the Langevin ensemble; its sampling uncertainty decreases with the number of independent replicas.
	
	The choice \(p\in\{1,2,4\}\) provides a low-order polynomial basis aligned with the linear drive and the quadratic and quartic terms of the onsite confinement. To state the physical meaning precisely, let
	\begin{equation}
		\begin{aligned}
			\bar{x}_i
			&=
			\mathbb{E}[x_i],
			&
			\sigma_i^2
			&=
			\mathbb{E}\!\left[
			(x_i-\bar{x}_i)^2
			\right],
			\\
			\mu^{\mathrm c}_{q,i}
			&=
			\mathbb{E}\!\left[
			(x_i-\bar{x}_i)^q
			\right].
		\end{aligned}
		\label{eq:central_moment_definitions}
	\end{equation}
	The second raw moment satisfies
	\begin{equation}
		m_{2,i}
		=
		\mathbb{E}[x_i^2]
		=
		\sigma_i^2+\bar{x}_i^2,
		\label{eq:raw_second_decomposition}
	\end{equation}
	and the fourth raw moment can be decomposed as
	\begin{equation}
		m_{4,i}
		=
		\mathbb{E}[x_i^4]
		=
		\mu^{\mathrm{c}}_{4,i}
		+
		4\bar{x}_i\mu^{\mathrm{c}}_{3,i}
		+
		6\bar{x}_i^2\sigma_i^2
		+
		\bar{x}_i^4.
		\label{eq:raw_fourth_decomposition}
	\end{equation}
	All moments in Eqs.~\eqref{eq:central_moment_definitions}--
	\eqref{eq:raw_fourth_decomposition} are evaluated at \(t_f\)
	and conditioned on the input \(\bm I\); these arguments are
	suppressed for compactness.
	
	Consequently, \(m_{2,i}\) is not a pure variance observable, and \(m_{4,i}\) is not a pure measure of kurtosis or non-Gaussian tail weight. Instead, they are raw polynomial observables that combine displacement and central-shape information. Their relevance here follows from their direct alignment with the quadratic and quartic powers appearing in the local dynamics. Higher raw moments could in principle be included, but their Monte Carlo estimators become increasingly sensitive to rare samples and generally exhibit larger finite-\(M\) variance.
	
	For a reservoir with input projection \(\bm K\), the local moment-resolved response representation is then defined as
	\begin{equation}
		\label{eq:moment_representation}
		\phi_{\bm{K}}(\bm{I})
		=
		\left[
		\bm{m}_1(\bm{I}),
		\bm{m}_2(\bm{I}),
		\bm{m}_4(\bm{I})
		\right]
		\in\mathbb{R}^{3N}.
	\end{equation}
	Equivalently, \(\bm m_p(\bm I)\) denotes the elementwise \(p\)-th raw moment vector of the finite-time response distribution. This construction maps the transient input-conditioned density \(P(\bm{x},t_f|\bm I)\) to a finite-dimensional moment-resolved response representation for subsequent classification. The mapping is intentionally incomplete: it retains selected nodewise polynomial moments but does not reconstruct the full probability density or the inter-node correlations \(\mathbb{E}[x_i x_j]\).
	
	\begin{figure*}[t!]
		\centering
		\includegraphics[width=0.48\linewidth]{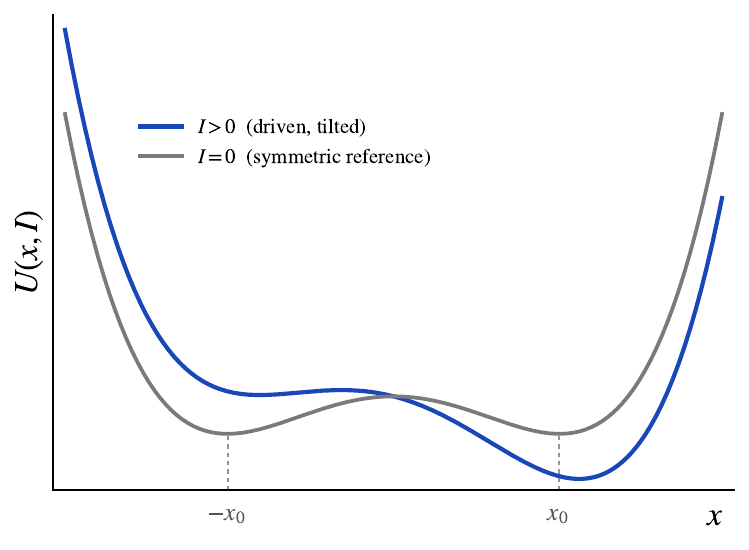}
		\hfill
		\includegraphics[width=0.48\linewidth]{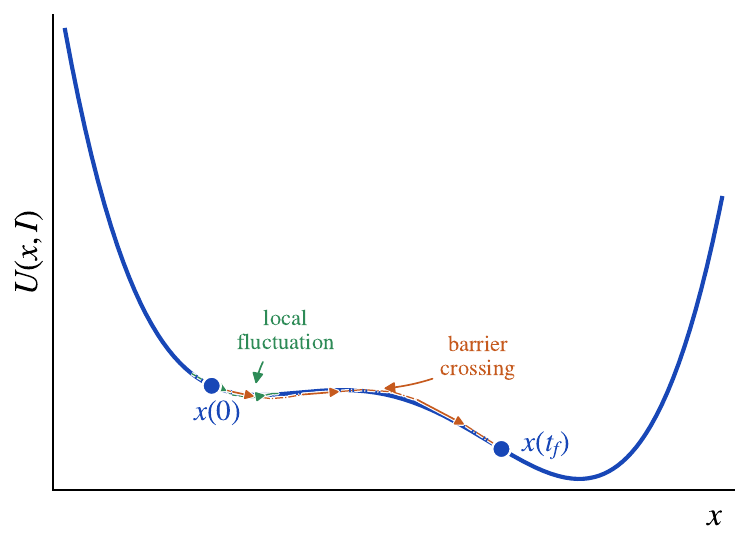}
		\caption{
			Physical mechanism of the thermodynamic neuron.
			(a) Driven quartic potential \(U(x,I)=J_2x^2+J_4x^4-Ix\) showing input-induced symmetry deformation relative to the symmetric reference case $I=0$.
			(b) Representative schematic overdamped Langevin paths on the driven nonlinear energy landscape, illustrating finite-time thermal excursions and nonlinear distribution reshaping.
		}
		\label{fig:thermodynamic_dynamics}
	\end{figure*}
	
	\begin{figure}[t!]
		\centering
		\includegraphics[width=0.95\linewidth]{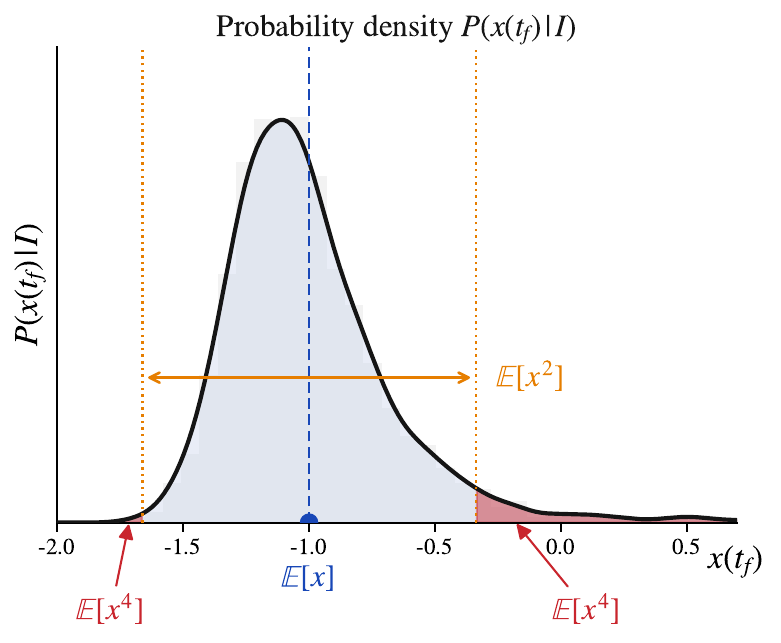}
		\caption{
			Moment-resolved thermodynamic response representation.
			Schematic finite-time probability density \(P(x(t_f)\mid I)\), whose raw moment channels
			\(\mathbb{E}[x]\), \(\mathbb{E}[x^2]\), and \(\mathbb{E}[x^4]\)
			encode mean displacement, fluctuation scale, and high-amplitude nonlinear excursions, respectively.
			These channels define the local moment-resolved response representation.
		}		
		\label{fig:moment_representation}
	\end{figure}

	\subsection{Nonequilibrium Trajectory Dynamics and Finite-Time Moment Readout}
	
	As shown in Fig.~\ref{fig:thermodynamic_dynamics}, the quartic-confined Langevin neuron converts an input drive into a tilted nonlinear energy landscape and finite-time stochastic trajectories. In the quartic-confined single-well response regime, the deterministic component of the force tends to relax the coordinate toward the global minimum, whereas the thermal bath continuously perturbs the trajectory away from purely deterministic descent. Consequently, different reset-sampled stochastic replicas undergo finite-time thermal excursions that reshape the nonequilibrium distribution within the observation time $t_f$, as illustrated schematically in Fig.~\ref{fig:thermodynamic_dynamics}(b).
	
	In this work, ``nonequilibrium'' is used in the operational finite-time sense: the system is reset to \(\bm{x}(0)=\bm 0\), driven by an input-dependent force, and read at a prescribed time \(t_f\) without assuming relaxation to a stationary distribution. The present classification experiment does not by itself quantify the distance between \(P(\bm{x},t_f|\bm I)\) and the corresponding long-time stationary distribution. Accordingly, the term finite-time nonequilibrium response refers to the reset-and-read protocol rather than to a separately established steady-state entropy-production regime. Let \(\mathcal{L}_{\bm I}^{\dagger}\) denote the Fokker--Planck operator associated with Eq.~\eqref{eq:fokker_planck} \cite{Risken1996}. For the reset-sampling protocol used in this work, the input-conditioned probability density can be written formally as
	\begin{equation}
		\label{eq:propagator}
		P(\bm{x},t_f|\bm I)
		=
		\exp\!\left(t_f\mathcal{L}_{\bm I}^{\dagger}\right)
		\delta(\bm{x}),
	\end{equation}
	where the delta distribution represents the reset initial condition \(\bm{x}(0)=\bm 0\). This expression emphasizes that the readout probes the finite-time propagator generated by the driven Langevin dynamics. It does not require the system to have relaxed to a stationary Boltzmann distribution at the observation time.
	
	Numerically, each stochastic replica is advanced using an Euler--Maruyama discretization of Eq.~\eqref{eq:langevin}~\cite{halidias_note_2008}:
	\begin{equation}
		\label{eq:euler_maruyama}
		x_{i,n+1}^{(m)}
		=
		x_{i,n}^{(m)}
		+
		\mu F_i(\bm{x}_{n}^{(m)},\bm I)\Delta t
		+
		\sqrt{2\mu k_B T\Delta t}\,
		\xi_{i,n}^{(m)},
	\end{equation}
	where \(\xi_{i,n}^{(m)}\sim\mathcal{N}(0,1)\) are independent Gaussian random variables across node index \(i\), time step \(n\), and replica index \(m\). The final-time samples
	\(\{\bm{x}^{(m)}(t_f;\bm I)\}_{m=1}^{M}\)
	provide a Monte Carlo representation of \(P(\bm{x},t_f|\bm I)\).
	
	For numerical safety, the updated state is projected after each Euler--Maruyama step onto the interval
	\begin{equation}
		x_{i,n+1}^{(m)}
		\leftarrow
		\Pi_{[-x_{\mathrm c},x_{\mathrm c}]}
		\!\left(
		x_{i,n+1}^{(m)}
		\right),
		\label{eq:hard_boundary}
	\end{equation}
	where
	\begin{equation}
		\Pi_{[-x_{\mathrm c},x_{\mathrm c}]}(y)
		=
		\min
		\left\{
		x_{\mathrm c},
		\max(-x_{\mathrm c},y)
		\right\}.
		\label{eq:hard_boundary_operator}
	\end{equation}
	where \(x_{\mathrm c}=4\) in the reported calculations. The implementation records the number of attempted updates that exceed this interval, so that the influence of the numerical boundary can be assessed independently from the classification result.

	The statistical readout of these finite-time trajectories is summarized separately in Fig.~\ref{fig:moment_representation}. Rather than compressing the distribution into the first-order response alone, the proposed moment-resolved thermodynamic response representation uses raw moment channels to sample distinct physical aspects of the nonequilibrium ensemble. 
	
	The first moment \(\mathbb{E}[x_i]\) measures the mean displacement response. The second raw moment \(\mathbb{E}[x_i^2]\) combines the variance with the squared mean displacement, while the fourth raw moment \(\mathbb{E}[x_i^4]\) combines fourth-order central-shape information with lower-order location and fluctuation contributions, as shown in Eqs.~\eqref{eq:raw_second_decomposition} and \eqref{eq:raw_fourth_decomposition}. These observables are therefore interpreted as progressively higher-order polynomial summaries of the finite-time response, rather than as pure variance and kurtosis measurements. The response vector in Eq.~\eqref{eq:moment_representation} retains information beyond the mean while remaining directly aligned with the polynomial structure of the onsite force~\cite{PhysRevLett.104.090601}.

	\section{Results and Discussion}

	\subsection{Experimental Protocol}
	
	We evaluate the proposed models on the standard MNIST handwritten-digit classification benchmark. The official training set contains \(60000\) samples and the official test set contains \(10000\) samples. The results reported here follow the fixed \(60000/10000\) protocol used during model development: reservoir construction and final readout fitting use the complete training set, and the reported accuracies are evaluated on the official test set.
	
	During an earlier exploratory stage, a \(2000\)-sample subset of the official test set was inspected for preliminary comparisons among feature combinations and fusion settings. Consequently, the reported \(10000\)-sample results should be interpreted as fixed-protocol reproduction results rather than as an untouched confirmatory estimate. For the final public reproduction, all numerical parameters,
	random seeds, reservoir definitions, feature channels, and
	readout hyperparameters will be fixed before regenerating the
	complete \(60000/10000\) results. No subsequent parameter
	adjustment will be made from the reproduced test predictions.
	
	Each \(28\times28\) grayscale image is flattened into a vector
	\(\bm I\in\mathbb{R}^{784}\), with pixel intensities scaled according to
	\begin{equation}
		I_k=\frac{u_k}{255},
		\qquad
		u_k\in\{0,1,\ldots,255\}.
		\label{eq:input_scaling}
	\end{equation}
	No pixelwise test-set statistics are used. Standardization is applied only to the extracted response features, using statistics computed from the training feature matrix.
	
	The finite-time Langevin simulations use
	\begin{equation}
		\begin{aligned}
			N&=256,
			&
			M&=128,
			&
			J_2&=1,
			&
			J_4&=1,
			\\
			\mu&=1,
			&
			k_BT&=0.1.
		\end{aligned}
		\label{eq:primary_dynamics_parameters}
	\end{equation}
	together with
	\begin{equation}
		\begin{aligned}
			\Delta t&=10^{-3},
			&
			t_f&=1,
			&
			N_t&=\frac{t_f}{\Delta t}=1000,
			\\
			\bm{x}(0)&=\bm 0.
		\end{aligned}
		\label{eq:integration_parameters}
	\end{equation}
	A hard numerical boundary \(x_{\mathrm c}=4\) is applied according to Eq.~\eqref{eq:hard_boundary_operator}. The primary stochastic trajectory seed is \(20260609\). Noise streams are generated independently across input samples, reservoir index, replica index, node index, and integration step. The same extracted feature realization is reused across nested moment-order ablations, so that the comparisons between
	\(\bm m_1\),
	\((\bm m_1,\bm m_2)\), and
	\((\bm m_1,\bm m_2,\bm m_4)\)
	are paired at the feature-extraction level.
	
	All reservoirs use \(N=256\) thermodynamic nodes. For each input image, \(M=128\) independent Langevin trajectories are sampled to estimate the raw response moments \(m_1\), \(m_2\), and \(m_4\). Thus, each single-reservoir moment-resolved response representation has dimension \(3N=768\). The full heterogeneous multi-reservoir response representation concatenates three reservoir blocks and has dimension \(9N=2304\).
	
	The three reservoirs are constructed using fixed but distinct initialization and surrogate-training histories. The recurrent matrices remain fixed during input-projection training. Reservoirs \(K_1\) and \(K_2\) use \(s_W=0.1\), whereas \(K_3\) uses \(s_W=0.2\). All input projections are initialized with \(s_K=1.2\), and the local biases are initialized with \(s_b=0.02\).
	
	The input projections are trained using deterministic local-response surrogates rather than by differentiating through the full stochastic Langevin simulation. For an input \(\bm I\), the local surrogate response \(x_i^\star\) is defined as the real stationary solution of
	\begin{equation}
		\begin{aligned}
		2J_2x_i^\star
		+
		4J_4(x_i^\star)^3
		&=
		h_i(\bm I), \\
		h_i(\bm I)
		&=
		b_i+\sum_{k=1}^{D}K_{ik}I_k.
		\end{aligned}
		\label{eq:local_surrogate_fixed_point}
	\end{equation}
	The mean-response surrogate uses
	\begin{equation}
		\bm\psi_{\mathrm{mean}}(\bm I)
		=
		\bm{x}^\star(\bm I),
		\label{eq:mean_surrogate}
	\end{equation}
	whereas the polynomial surrogate uses
	\begin{equation}
		\bm\psi_{\mathrm{poly}}(\bm I)
		=
		\left[
		\bm{x}^\star(\bm I),
		(\bm{x}^\star(\bm I))^{\odot2},
		(\bm{x}^\star(\bm I))^{\odot4}
		\right].
		\label{eq:polynomial_surrogate}
	\end{equation}
	These surrogates do not include thermal noise, finite-time integration, or recurrent coupling and are used only to initialize and optimize the input projections. All classification features reported in Tables~\ref{tab:ablation}--\ref{tab:moment_order_ablation} are subsequently regenerated using the full finite-time coupled Langevin dynamics of Eq.~\eqref{eq:langevin}.
	
	Reservoir \(K_1\) is trained directly with the polynomial surrogate for \(60\) epochs using learning rate \(10^{-3}\) and seed \(20260611\). Reservoir \(K_2\) is first trained with the mean-response surrogate for \(100\) epochs using learning rate \(3\times10^{-3}\), and is then refined for \(60\) epochs with the polynomial surrogate using learning rate \(10^{-3}\) and seed \(20260612\). Reservoir \(K_3\) is independently initialized and trained with the polynomial surrogate for \(80\) epochs using learning rate \(10^{-3}\), \(s_W=0.2\), and seed \(20260613\). The surrogate-training batch size is \(128\), and the regularization coefficient on the trained input projection is \(10^{-5}\).
	
	The main comparison includes single-reservoir representations,
	pairwise concatenated representations, and the proposed
	three-reservoir feature-level model. Equal-weight
	decision-level logit averaging is included as a fixed diagnostic
	baseline. This comparison evaluates one simple form of late
	fusion and is not intended as an exhaustive comparison with
	calibrated, validation-weighted, or learned stacking methods. We denote a single-reservoir moment-resolved representation by
	SR-MRR, a pairwise concatenated representation by PR-MRR, the
	heterogeneous multi-reservoir feature-level model by HMR-FL,
	and the equal-weight decision-level logit-fusion baseline by
	HMR-LF.
	
	\begin{table*}[t]
		\centering
		\caption{
			Ablation study of moment-resolved response representation models on MNIST.
			The readout is trained on the full MNIST training set of 60000 samples and evaluated on the official MNIST test set of 10000 samples.
			Each single reservoir contributes \(3N=768\) moment features with \(N=256\).
			Pairwise reservoirs use \(1536\) features, and the full three-reservoir feature-level fusion uses \(2304\) features.  
		}
		\label{tab:ablation}
		\begin{ruledtabular}
			\begin{tabular}{lccc}
				Method & Feature dim. & Test correct & Test accuracy \\
				\hline
				SR-MRR \((K_1)\) & 768 & \(9651/10000\) & \(96.51\%\) \\
				SR-MRR \((K_2)\) & 768 & \(9682/10000\) & \(96.82\%\) \\
				SR-MRR \((K_3)\) & 768 & \(9603/10000\) & \(96.03\%\) \\
				\hline
				PR-MRR \((K_1,K_2)\) & 1536 & \(9658/10000\) & \(96.58\%\) \\
				PR-MRR \((K_1,K_3)\) & 1536 & \(9683/10000\) & \(96.83\%\) \\
				PR-MRR \((K_2,K_3)\) & 1536 & \(9680/10000\) & \(96.80\%\) \\
				\hline
				HMR-FL \((K_1,K_2,K_3)\) & 2304 & \(9695/10000\) & \(96.95\%\) \\
			\end{tabular}
		\end{ruledtabular}
	\end{table*}
	
	\begin{table*}[t]
		\centering
		\caption{
			Comparison of the strongest single-reservoir baseline, equal-weight decision-level logits fusion, and the proposed feature-level fusion.
			The late-fusion baseline first compresses each \(768\)-dimensional reservoir response into independent logits before averaging.
			The proposed feature-level fusion trains a single readout on the uncompressed \(2304\)-dimensional joint moment-resolved response representation.
		}
		\label{tab:fusion_comparison}
		\begin{ruledtabular}
			\begin{tabular}{lcccc}
				Method & Feature/readout space & Fusion rule & Test correct & Test accuracy \\
				\hline
				Best SR-MRR & \(768\) & \(K_2\) only & \(9682/10000\) & \(96.82\%\) \\
				HMR-LF (equal) & \(3\times 10\) logits & \((\bm{z}_1+\bm{z}_2+\bm{z}_3)/3\) & \(9684/10000\) & \(96.84\%\) \\
				HMR-FL (proposed) & \(2304\) features & \([\phi_{\bm K_1},\phi_{\bm K_2},\phi_{\bm K_3}]\) & \(9695/10000\) & \(96.95\%\) \\
			\end{tabular}
		\end{ruledtabular}
	\end{table*}

	\begin{table*}[t]
		\centering
		\caption{
			Wrong-set overlap analysis on the full MNIST test set.
			For a model \(A\), let \(\mathcal{E}_A\) denote the set of test samples misclassified by \(A\).
			The Jaccard wrong-set overlap is
			\(J(A,B)=|\mathcal{E}_A\cap\mathcal{E}_B|/|\mathcal{E}_A\cup\mathcal{E}_B|\).
			Lower Jaccard values indicate lower overlap between the two observed error sets. This label-level statistic does not by itself establish feature-space decorrelation or collinearity.
		}
		\label{tab:jaccard}
		\begin{ruledtabular}
			\begin{tabular}{lccccc}
				Model pair & \(|\mathcal{E}_A|\) & \(|\mathcal{E}_B|\) &
				\(|\mathcal{E}_A\cap\mathcal{E}_B|\) &
				\(|\mathcal{E}_A\cup\mathcal{E}_B|\) &
				Jaccard \\
				\hline
				\(K_1\) vs. \(K_2\) & 349 & 318 & 275 & 392 & 0.7015 \\
				\(K_1\) vs. \(K_3\) & 349 & 397 & 275 & 471 & 0.5839 \\
				\(K_2\) vs. \(K_3\) & 318 & 397 & 250 & 465 & 0.5376 \\
				\hline
				\(K_1\) vs. \(K_1+K_2+K_3\) & 349 & 305 & 258 & 396 & 0.6515 \\
				\(K_2\) vs. \(K_1+K_2+K_3\) & 318 & 305 & 250 & 373 & 0.6702 \\
				\(K_3\) vs. \(K_1+K_2+K_3\) & 397 & 305 & 254 & 448 & 0.5670 \\
			\end{tabular}
		\end{ruledtabular}
	\end{table*}
	
	The full HMR-FL
	 model reduces the number of full-test errors from 349, 318, and 397 for \(K_1\), \(K_2\), and \(K_3\), respectively, to 305.
	
	\subsection{Error-Set Complementarity in Heterogeneous Reservoirs}
	
	The individual and pairwise results are summarized in Table~\ref{tab:ablation}. Among the single-reservoir models, \(K_2\) gives the highest observed test accuracy, \(96.82\%\). Concatenating response features does not produce a monotonic improvement: the \(K_1+K_2\) model reaches \(96.58\%\), below the standalone \(K_2\) result. This observation shows that increasing feature dimension alone is not sufficient to guarantee improved test performance under the fixed readout protocol.
	
	The wrong-set Jaccard coefficient provides a descriptive measure of overlap between the samples misclassified by two models~\cite{648054.743935,annurev:/content/journals/10.1146/annurev-conmatphys-031119-050745,Hansen1990NeuralNE,LUKOSEVICIUS2009127}. The \(K_1\)--\(K_2\) pair has the largest observed wrong-set overlap, \(J=0.7015\), whereas the \(K_2\)--\(K_3\) pair has the lowest, \(J=0.5376\). These values indicate differences in classification-error overlap, but they do not directly measure feature-space correlation, canonical angles, or linear collinearity. The reduced \(K_1+K_2\) accuracy is therefore consistent with partially redundant classification behavior, but the Jaccard analysis alone does not establish redundancy as its cause. Feature dimensionality, optimization, and regularization may also contribute.
	
	Relative to \(K_2\), the full \(K_1+K_2+K_3\) feature model corrects \(68\) samples that \(K_2\) misclassifies and introduces \(55\) errors on samples that \(K_2\) classifies correctly. The exact two-sided McNemar test on these discordant pairs gives \(p\simeq0.279\). The \(0.13\)-percentage-point difference is therefore reported as the best observed accuracy among the compared configurations, not as a statistically established improvement over \(K_2\).
	
	Although \(K_3\) is the weakest standalone model, its error set overlaps less with that of \(K_2\) than does the error set of \(K_1\). The full feature-level model reaches \(96.95\%\), suggesting that a weaker reservoir may still provide useful coordinates when its classification errors differ from those of stronger reservoirs. This relationship is descriptive rather than causal: wrong-set overlap does not demonstrate that the corresponding feature blocks are statistically decorrelated.

	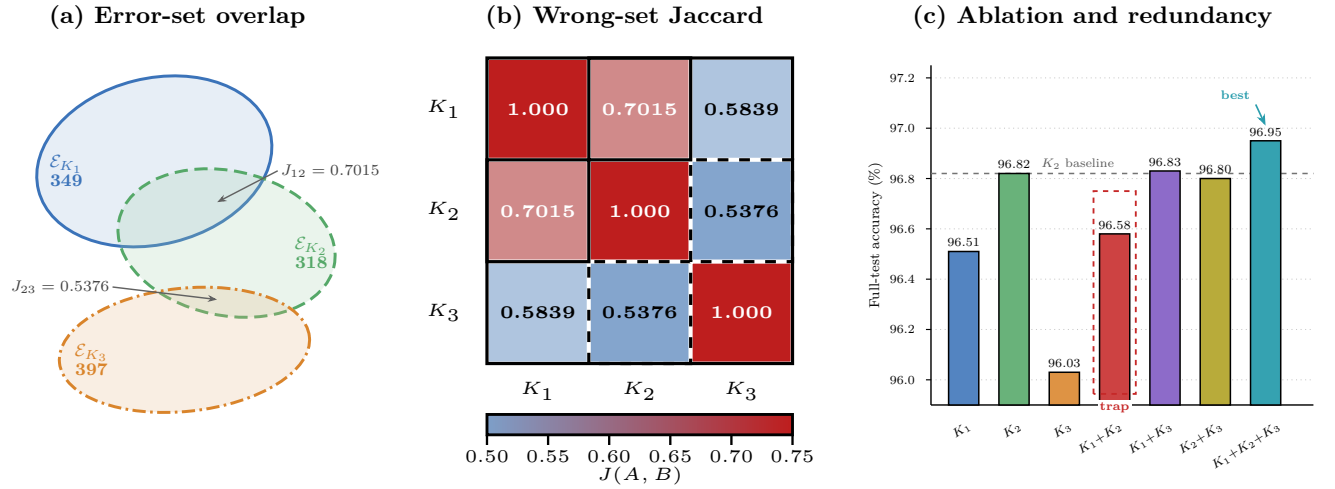
\begin{figure*}[t!]
		\centering
		
		\begin{minipage}[t][7.0cm][t]{0.31\linewidth}
			\centering
			{\small\bfseries (a) Error-set overlap}\par\vspace{0.6ex}
			
			\resizebox{\linewidth}{!}{%
				\begin{tikzpicture}[
					>=Stealth,
					every node/.style={font=\footnotesize},
					lab/.style={font=\footnotesize\bfseries},
					ann/.style={draw=black!62, -{Stealth[scale=0.72]}, line width=0.65pt},
					jlab/.style={
						font=\scriptsize\bfseries,
						text=black!75,
						fill=white,
						fill opacity=0.82,
						text opacity=1,
						inner sep=1.2pt,
						rounded corners=1pt
					}
					]
					
					\coordinate (EoneC)   at (0.00,  0.65);
					\coordinate (EtwoC)   at (1.15, -0.65);
					\coordinate (EthreeC) at (0.48, -2.35);
					
					\draw[
					draw=cKOne,
					line width=1.35pt,
					fill=cKOne!55,
					fill opacity=0.22,
					rotate around={14:(EoneC)}
					]
					(EoneC) ellipse (1.90 and 1.32);
					
					\draw[
					draw=cKTwo,
					line width=1.35pt,
					dash pattern=on 6pt off 3pt,
					fill=cKTwo!55,
					fill opacity=0.22,
					rotate around={-12:(EtwoC)}
					]
					(EtwoC) ellipse (1.75 and 1.15);
					
					\draw[
					draw=cKThree,
					line width=1.35pt,
					dash pattern=on 6pt off 2pt on 1.2pt off 2pt,
					fill=cKThree!60,
					fill opacity=0.22,
					rotate around={6:(EthreeC)}
					]
					(EthreeC) ellipse (2.0 and 0.98);
					
					\node[lab,text=cKOne,align=center] (lkone) at (-1.40,0.50)
					{$\mathcal{E}_{K_1}$\\[-1pt]349};
					
					\node[lab,text=cKTwo,align=center] (lktwo) at (2.50,-0.80)
					{$\mathcal{E}_{K_2}$\\[-1pt]318};
					
					\node[lab,text=cKThree,align=center] (lkthree) at (-1.00,-2.50)
					{$\mathcal{E}_{K_3}$\\[-1pt]397};
					
					\node[jlab] (j12) at (2.80,0.50) {$J_{12}=0.7015$};
					\draw[ann] (j12.west) -- (1.00,0.00);
					
					\node[jlab] (j23) at (-1.50,-1.35) {$J_{23}=0.5376$};
					\draw[ann] (j23.east) -- (1.00,-1.55);
					
					\useasboundingbox (-2.95,-3.70) rectangle (3.45,2.62);
					
				\end{tikzpicture}
			}
		\end{minipage}
		\hfill
		\begin{minipage}[t][7.0cm][t]{0.31\linewidth}
			\centering
			{\small\bfseries (b) Wrong-set Jaccard}\par\vspace{0.6ex}
			
			\resizebox{\linewidth}{!}{%
				\begin{tikzpicture}[
					every node/.style={font=\footnotesize},
					heatnum/.style={font=\tiny\bfseries},
					ticklab/.style={font=\tiny}
					]
					
					\newcommand{\hmcell}[5]{%
						\filldraw[#3, draw=white, line width=1.1pt] (#1,#2) rectangle ++(1,1);
						\node[heatnum,#5] at (#1+0.5,#2+0.5) {#4};
					}
					
					\hmcell{0}{2}{cHeatHigh}{1.000}{white}
					\hmcell{1}{2}{cHeatMid}{0.7015}{white}
					\hmcell{2}{2}{cHeatLow!55}{0.5839}{black}
					
					\hmcell{0}{1}{cHeatMid}{0.7015}{white}
					\hmcell{1}{1}{cHeatHigh}{1.000}{white}
					\hmcell{2}{1}{cHeatLow!85}{0.5376}{black}
					
					\hmcell{0}{0}{cHeatLow!55}{0.5839}{black}
					\hmcell{1}{0}{cHeatLow!85}{0.5376}{black}
					\hmcell{2}{0}{cHeatHigh}{1.000}{white}
					
					\draw[black, line width=0.9pt] (0,0) rectangle (3,3);
					
					\draw[black, line width=0.9pt] (1,2) rectangle (2,3);
					\draw[black, line width=0.9pt] (0,1) rectangle (1,2);
					
					\draw[black, dashed, line width=0.9pt] (2,1) rectangle (3,2);
					\draw[black, dashed, line width=0.9pt] (1,0) rectangle (2,1);
					
					\node[ticklab,left=4pt]  at (0,2.5) {$K_1$};
					\node[ticklab,left=4pt]  at (0,1.5) {$K_2$};
					\node[ticklab,left=4pt]  at (0,0.5) {$K_3$};
					
					\node[ticklab,below=2pt] at (0.5,0) {$K_1$};
					\node[ticklab,below=2pt] at (1.5,0) {$K_2$};
					\node[ticklab,below=2pt] at (2.5,0) {$K_3$};
					
					\shade[
					left color=cHeatLow!90,
					middle color=white,
					right color=cHeatHigh
					] (0.00,-0.70) rectangle (3.00,-0.50);
					
					\draw[black, line width=0.8pt] (0.00,-0.70) rectangle (3.00,-0.50);
					
					\foreach \x/\txt in {
						0.00/0.50,
						0.60/0.55,
						1.20/0.60,
						1.80/0.65,
						2.40/0.70,
						3.00/0.75
					}{
						\draw[black, line width=0.65pt] (\x,-0.70) -- (\x,-0.80);
						\node[font=\tiny, below=1pt] at (\x,-0.70) {\txt};
					}
					
					\node[font=\tiny] at (1.5,-1.10) {$J(A,B)$};
					
					\useasboundingbox (-0.52,-1.35) rectangle (3.28,3.20);
					
				\end{tikzpicture}
			}
		\end{minipage}
		\hfill
		\begin{minipage}[t][7.0cm][t]{0.34\linewidth}
			\centering
			{\small\bfseries (c) Ablation and redundancy}\par\vspace{0.6ex}
			
			\resizebox{\linewidth}{!}{%
				\begin{tikzpicture}[
					every node/.style={font=\footnotesize},
					val/.style={font=\scriptsize},
					ticklab/.style={font=\scriptsize},
					>=Stealth
					]
					
					%
					\def\ymin{95.9}
					\def\yscale{5.0}
					
					\pgfmathsetmacro{\yaxismax}{((97.25)-\ymin)*\yscale}
					\pgfmathsetmacro{\ylabely}{0.5*\yaxismax}
					\pgfmathsetmacro{\besty}{((96.95)-\ymin)*\yscale}
					\pgfmathsetmacro{\basey}{((96.82)-\ymin)*\yscale}
					\pgfmathsetmacro{\trapTop}{((96.75)-\ymin)*\yscale}
					
					\newcommand{\barc}[4]{%
						\pgfmathsetmacro{\yb}{((#2)-\ymin)*\yscale}
						\filldraw[fill=#3, draw=black, line width=0.85pt]
						(#1-0.30,0) rectangle (#1+0.30,\yb);
						\node[val] at (#1,\yb+0.18) {#4};
					}
					
					\foreach \v in {96.0,96.2,96.4,96.6,96.8,97.0,97.2}{
						\pgfmathsetmacro{\gy}{((\v)-\ymin)*\yscale}
						\draw[black!22, dotted, line width=0.58pt] (-0.65,\gy) -- (6.95,\gy);
						\draw[black, line width=0.65pt] (-0.65,\gy) -- (-0.75,\gy);
						\node[ticklab,left=3pt] at (-0.75,\gy) {\v};
					}
					
					\draw[black, line width=0.8pt] (-0.65,0) -- (6.95,0);
					\draw[black, line width=0.8pt] (-0.65,0) -- (-0.65,\yaxismax);
					
					\draw[black!55, dashed, line width=0.85pt] (-0.65,\basey) -- (6.95,\basey);
					
					\node[
					font=\scriptsize,
					text=black!65,
					anchor=west,
					fill=white,
					inner sep=1.2pt
					] at (1.50,\basey+0.20) {$K_2$ baseline};
					
					\barc{0}{96.51}{cKOne!88}{96.51}
					\barc{1}{96.82}{cKTwo!88}{96.82}
					\barc{2}{96.03}{cKThree!88}{96.03}
					\barc{3}{96.58}{cKOneTwo!88}{96.58}
					\barc{4}{96.83}{cKOneThree!88}{96.83}
					\barc{5}{96.80}{cKTwoThree!92}{96.80}
					\barc{6}{96.95}{cKAll!95}{96.95}
					
					\node[ticklab,rotate=32,anchor=north east] at (0.10,-0.16) {$K_1$};
					\node[ticklab,rotate=32,anchor=north east] at (1.10,-0.16) {$K_2$};
					\node[ticklab,rotate=32,anchor=north east] at (2.10,-0.16) {$K_3$};
					\node[ticklab,rotate=32,anchor=north east] at (3.10,-0.16) {$K_1{+}K_2$};
					\node[ticklab,rotate=32,anchor=north east] at (4.10,-0.16) {$K_1{+}K_3$};
					\node[ticklab,rotate=32,anchor=north east] at (5.10,-0.16) {$K_2{+}K_3$};
					\node[ticklab,rotate=32,anchor=north east] at (6.25,-0.16) {$K_1{+}K_2{+}K_3$};
					
					\draw[cKOneTwo, dashed, line width=1.05pt]
					(2.58,0.22) rectangle (3.42,\trapTop);
					
					\node[
					font=\scriptsize\bfseries,
					text=cKOneTwo,
					fill=white,
					inner sep=1pt
					] at (3.0,-0.05) {trap};
					
					\node[
					font=\scriptsize\bfseries,
					text=cKAll
					] (bestlab) at (5.40,\besty+0.92) {best};
					
					\draw[-{Stealth[scale=0.9]}, cKAll, line width=0.95pt]
					(bestlab.south east) -- (6.00,\besty+0.30);
					
					\node[rotate=90, font=\footnotesize]
					at (-1.75,\ylabely) {Full-test accuracy (\%)};
					
					\useasboundingbox (-1.75,-0.90) rectangle (7.05,\yaxismax+0.55);
					
				\end{tikzpicture}
			}
		\end{minipage}
		
		\caption{
			Error-mode decorrelation analysis on the MNIST full test set.
			(a) Schematic error sets \(\mathcal{E}_{K_r}\); numbers denote misclassified test samples, and \(J_{ij}\) denotes the Jaccard overlap between wrong sets.
			(b) Pairwise wrong-set Jaccard overlap matrix, with solid boxes marking the highly overlapping \(K_1\)--\(K_2\) pair and dashed boxes marking the more decorrelated \(K_2\)--\(K_3\) pair.
			(c) Ablation accuracy for individual, pairwise, and full three-reservoir response representations.
			The \(K_1+K_2\) result illustrates the performance degradation under concatenation, whereas full feature-level fusion gives the best observed test accuracy among the compared configurations.
		}
		
		\label{fig:error_decorrelation}
	\end{figure*}
	
	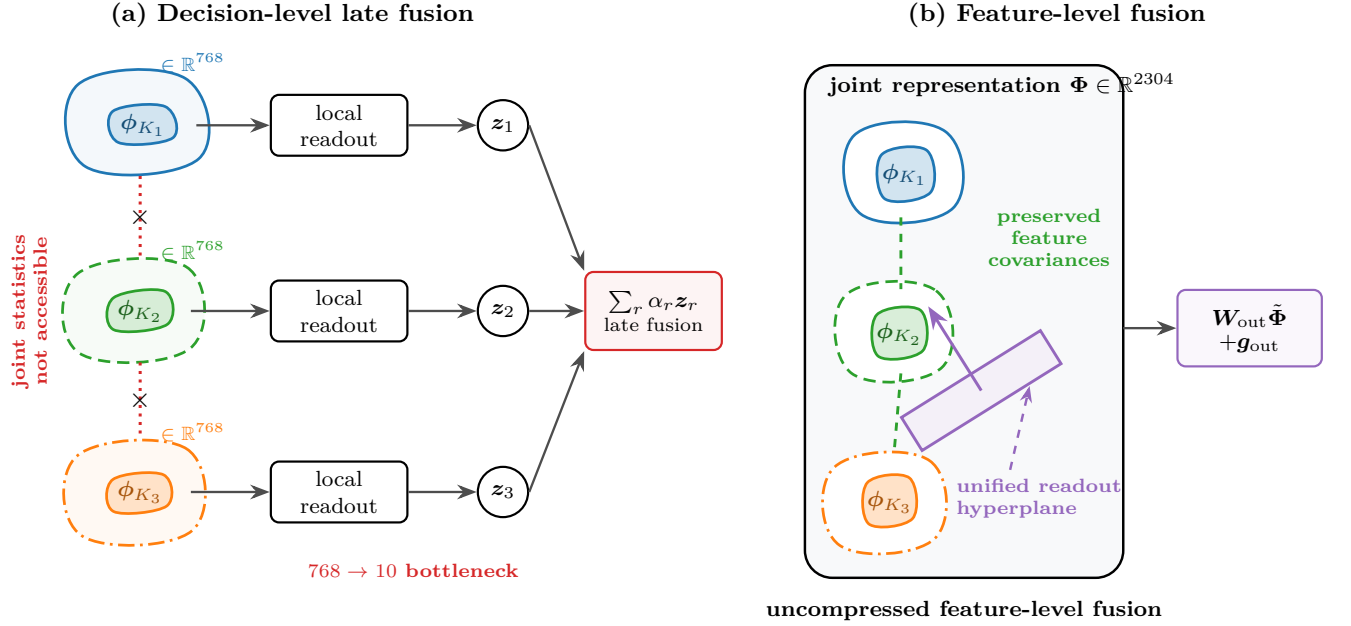
\begin{figure*}[htpb]
		\centering
		\resizebox{0.98\linewidth}{!}{%
			\begin{tikzpicture}[
				box/.style={
					rectangle,
					draw=black,
					line width=0.8pt,
					minimum width=1.75cm,
					minimum height=0.78cm,
					rounded corners=3pt,
					align=center,
					fill=white,
					font=\footnotesize
				},
				nodeCircle/.style={
					circle,
					draw=black,
					line width=0.8pt,
					minimum size=0.62cm,
					align=center,
					fill=white,
					font=\footnotesize
				},
				arrowLine/.style={
					-{Stealth[scale=1.05]},
					line width=0.85pt,
					draw=black!70
				},
				cutLine/.style={
					draw=compRed,
					line width=1.1pt,
					dotted
				},
				manifoldLabel/.style={
					font=\footnotesize\bfseries
				},
				dimLabel/.style={
					font=\scriptsize\bfseries
				}
				]
				
				\node[align=center, font=\bfseries] at (2.65,4.10)
				{(a) Decision-level late fusion};
				
				\draw[
				line width=1.0pt,
				compBlue,
				fill=compBlue!5
				]
				plot[smooth cycle, tension=0.70]
				coordinates {(-0.10,3.20) (1.28,3.28) (1.45,2.25) (0.00,2.15)};
				
				\draw[
				line width=1.0pt,
				compBlue,
				fill=compBlue!20
				]
				plot[smooth cycle, tension=0.62]
				coordinates {(0.35,2.86) (1.02,2.92) (1.12,2.52) (0.42,2.46)};
				
				\node[compBlue!65!black, manifoldLabel]
				at (0.74,2.68) {$\bm{\phi}_{K_1}$};
				\node[compBlue, dimLabel, anchor=south west]
				at (0.85, 3.25) {$\in \mathbb{R}^{768}$};
				
				\draw[
				line width=1.0pt,
				compGreen,
				dash pattern=on 4.5pt off 2.5pt,
				fill=compGreen!4
				]
				plot[smooth cycle, tension=0.70]
				coordinates {(-0.10,0.78) (1.30,0.86) (1.34,-0.18) (-0.08,-0.24)};
				
				\draw[
				line width=1.0pt,
				compGreen,
				fill=compGreen!19
				]
				plot[smooth cycle, tension=0.62]
				coordinates {(0.36,0.47) (1.02,0.55) (1.05,0.13) (0.36,0.06)};
				
				\node[compGreen!60!black, manifoldLabel]
				at (0.70,0.30) {$\bm{\phi}_{K_2}$};
				\node[compGreen, dimLabel, anchor=south west]
				at (0.85, 0.84) {$\in \mathbb{R}^{768}$};
				
				\draw[
				line width=1.0pt,
				compOrange,
				dash pattern=on 6pt off 2pt on 1pt off 2pt,
				fill=compOrange!5
				]
				plot[smooth cycle, tension=0.70]
				coordinates {(-0.10,-1.55) (1.30,-1.47) (1.36,-2.52) (-0.05,-2.58)};
				
				\draw[
				line width=1.0pt,
				compOrange,
				fill=compOrange!23
				]
				plot[smooth cycle, tension=0.62]
				coordinates {(0.35,-1.86) (1.00,-1.80) (1.04,-2.20) (0.36,-2.27)};
				
				\node[compOrange!65!black, manifoldLabel]
				at (0.70,-2.02) {$\bm{\phi}_{K_3}$};
				\node[compOrange, dimLabel, anchor=south west]
				at (0.85, -1.49) {$\in \mathbb{R}^{768}$};
				
				\draw[cutLine] (0.70,2.03) -- (0.70,0.98)
				node[midway, scale=1.10, font=\bfseries] {$\times$};
				\draw[cutLine] (0.70,-0.36) -- (0.70,-1.32)
				node[midway, scale=1.10, font=\bfseries] {$\times$};
				
				\node[
				compRed,
				rotate=90,
				font=\scriptsize\bfseries,
				align=center
				] at (-0.72,0.30) {joint statistics\\not accessible};
				
				\node[box] (lr1) at (3.25,2.68) {local\\readout};
				\node[box] (lr2) at (3.25,0.30) {local\\readout};
				\node[box] (lr3) at (3.25,-2.02) {local\\readout};
				
				\draw[arrowLine] (1.42,2.68) -- (lr1.west);
				\draw[arrowLine] (1.35,0.30) -- (lr2.west);
				\draw[arrowLine] (1.35,-2.02) -- (lr3.west);
				
				\node[nodeCircle] (z1) at (5.35,2.68) {$\bm{z}_1$};
				\node[nodeCircle] (z2) at (5.35,0.30) {$\bm{z}_2$};
				\node[nodeCircle] (z3) at (5.35,-2.02) {$\bm{z}_3$};
				
				\draw[arrowLine] (lr1.east) -- (z1.west);
				\draw[arrowLine] (lr2.east) -- (z2.west);
				\draw[arrowLine] (lr3.east) -- (z3.west);
				
				\node[
				compRed,
				font=\scriptsize\bfseries
				] at (4.20,-3.02) {$768\rightarrow 10$ bottleneck};
				
				\node[
				box,
				draw=compRed,
				fill=compRed!5,
				minimum width=1.75cm,
				minimum height=1.00cm
				] (lf) at (7.28,0.30)
				{$\sum_r \alpha_r\bm{z}_r$\\[-1pt]\scriptsize late fusion};
				
				\draw[arrowLine] (z1.east) -- (lf.north west);
				\draw[arrowLine] (z2.east) -- (lf.west);
				\draw[arrowLine] (z3.east) -- (lf.south west);
				
				\begin{scope}[shift={(9.40,0.0)}]
					
					\node[align=center, font=\bfseries] at (3.05,4.10)
					{(b) Feature-level fusion};
					
					\draw[
					line width=0.9pt,
					black,
					fill=bgGray,
					rounded corners=10pt
					] (-0.18,-3.12) rectangle (3.90,3.45);
					
					\node[
					anchor=west,
					font=\bfseries\footnotesize
					] at (0.02,3.20)
					{joint representation $\bm{\Phi}\in\mathbb{R}^{2304}$};
					
					\coordinate (planeC) at (2.08,-0.72);
					\coordinate (normalTip) at (1.40,0.36);
					\coordinate (calloutTarget) at ($(planeC)+(0.62,0.06)$);
					
					\begin{scope}[shift={(planeC)}, rotate=32]
						\draw[
						line width=1.05pt,
						draw=compPurple,
						fill=compPurple!10
						] (-1.05,-0.25) rectangle (1.05,0.25);
					\end{scope}
					
					\draw[
					line width=1.0pt,
					compBlue,
					fill=white
					]
					plot[smooth cycle, tension=0.72]
					coordinates {(0.45,2.55) (1.62,2.60) (1.72,1.62) (0.54,1.54)};
					
					\draw[
					line width=1.0pt,
					compBlue,
					fill=compBlue!20
					]
					plot[smooth cycle, tension=0.62]
					coordinates {(0.80,2.30) (1.38,2.36) (1.43,1.82) (0.84,1.74)}; 
					
					\node[compBlue!65!black, manifoldLabel]
					at (1.10,2.05) {$\bm{\phi}_{K_1}$};
					
					\draw[
					line width=1.0pt,
					compGreen,
					dash pattern=on 4.5pt off 2.5pt,
					fill=white
					]
					plot[smooth cycle, tension=0.72]
					coordinates {(0.34,0.50) (1.50,0.56) (1.58,-0.42) (0.42,-0.50)};
					
					\draw[
					line width=1.0pt,
					compGreen,
					fill=compGreen!18
					]
					plot[smooth cycle, tension=0.62]
					coordinates {(0.74,0.25) (1.30,0.31) (1.33,-0.21) (0.76,-0.28)}; 
					
					\node[compGreen!60!black, manifoldLabel]
					at (1.03,0.01) {$\bm{\phi}_{K_2}$};
					
					\draw[
					line width=1.0pt,
					compOrange,
					dash pattern=on 6pt off 2pt on 1pt off 2pt,
					fill=white
					]
					plot[smooth cycle, tension=0.72]
					coordinates {(0.20,-1.70) (1.42,-1.62) (1.48,-2.62) (0.26,-2.70)};
					
					\draw[
					line width=1.0pt,
					compOrange,
					fill=compOrange!23
					]
					plot[smooth cycle, tension=0.62]
					coordinates {(0.62,-1.90) (1.16,-1.84) (1.20,-2.36) (0.64,-2.43)}; 
					
					\node[compOrange!65!black, manifoldLabel]
					at (0.91,-2.15) {$\bm{\phi}_{K_3}$};
					
					\draw[
					line width=1.05pt,
					compGreen,
					dashed
					] (1.05,1.50) -- (1.05,0.58);
					
					\draw[
					line width=1.05pt,
					compGreen,
					dashed
					] (1.05,-0.50) -- (0.96,-1.55);
					
					\node[
					compGreen,
					font=\scriptsize\bfseries,
					align=center
					] at (2.95,1.24)
					{preserved\\feature\\covariances};
					
					\draw[
					-{Stealth[scale=1.05]},
					compPurple,
					line width=1.10pt
					] (planeC) -- (normalTip);
					
					\node[
					compPurple,
					font=\scriptsize\bfseries,
					align=left,
					anchor=west
					] (hlLabel) at (1.65,-2.10)
					{unified readout\\hyperplane};
					
					\draw[
					-{Stealth[scale=0.85]},
					compPurple,
					line width=0.90pt,
					dashed
					] ([xshift=0.8cm]hlLabel.north west) -- (calloutTarget);
					
					\node[
					box,
					draw=compPurple,
					fill=compPurple!5,
					minimum width=1.85cm,
					minimum height=0.98cm
					] (um) at (5.52,0.08)
					{$\bm{W}_{\mathrm{out}}\tilde{\bm{\Phi}}$\\[-1pt]$+\bm{g}_{\mathrm{out}}$};
					
					\draw[arrowLine] (3.90,0.08) -- (um.west);
					
					\node[
					anchor=north,
					font=\bfseries\footnotesize
					] at (1.86,-3.30)
					{uncompressed feature-level fusion};
					
				\end{scope}
				
			\end{tikzpicture}
		}
		
		\caption{
			Geometric schematic of information mapping in decision-level and feature-level fusion.
			(a) In decision-level late fusion, each reservoir response block
			\(\bm{\phi}_{K_r}\in\mathbb{R}^{768}\) is first compressed by an independent local readout into a
			\(10\)-dimensional logit vector \(\bm{z}_r\). This early bottleneck constrains each reservoir to an
			independently optimized representation, preventing the final fusion layer from leveraging the full
			uncompressed feature geometry.
			(b) In feature-level fusion, the three response blocks are concatenated into the full thermodynamic
			response representation \(\bm{\Phi}\in\mathbb{R}^{2304}\). A unified readout hyperplane acts directly on the
			uncompressed joint feature manifold, jointly optimizing all class-feature weights before reservoir-wise 
			compression and enabling collective discrimination.
		}
		\label{fig:fusion_mechanics}
	\end{figure*}
	\subsection{Moment-Order Ablation of Thermodynamic Response Representations}
	
	To isolate the physical contribution of the different response channels, we perform a formal moment-order ablation, as shown in Table~\ref{tab:moment_order_ablation}, using the same computational readout pipeline, optimization schedule, and \(L_2\) penalty as the main HMR-FL
	 result.
	
	\begin{table}[t]
		\centering
		\caption{
			Formal moment-order ablation of the heterogeneous multi-reservoir moment-resolved response representation on the MNIST full test set.
			All rows use the same numerical readout pipeline as the main HMR-FL
			 result:
			150 epochs, batch size 128, learning rate \(0.05\), and \(L_2=10^{-3}\).
		}
		\label{tab:moment_order_ablation}
		\begin{ruledtabular}
			\begin{tabular}{lccc}
				Feature channels & Feature dim. & Test correct & Test accuracy \\
				\hline
				\(\bm m_1\) & 768 & \(9377/10000\) & \(93.77\%\) \\
				\(\bm m_1,\bm m_2\) & 1536 & \(9682/10000\) & \(96.82\%\) \\
				\(\bm m_1,\bm m_2,\bm m_4\) & 2304 & \(9695/10000\) & \(96.95\%\) \\
			\end{tabular}
		\end{ruledtabular}
	\end{table}
	
	The mean-only representation reaches \(93.77\%\). Adding the second raw moment increases the observed accuracy to \(96.82\%\), while adding the fourth raw moment gives \(96.95\%\). Because \(m_2=\sigma^2+\bar{x}^2\), the improvement obtained by adding \(\bm m_2\) cannot be attributed exclusively to fluctuation variance: the channel also contains nonlinear information about the squared mean displacement. Likewise, \(\bm m_4\) combines fourth-order central-shape information with lower-order location and fluctuation contributions.
	
	The difference between
	\((\bm m_1,\bm m_2)\) and
	\((\bm m_1,\bm m_2,\bm m_4)\)
	is \(13\) correctly classified samples, or \(0.13\) percentage points, in the reported run. This is a small additional observed improvement and is not described here as a statistically established or systematic effect. The ablation supports the usefulness of progressively higher-order raw polynomial observables under the fixed protocol, but further trajectory and readout repetitions are needed to distinguish a robust fourth-order contribution from finite-sampling variability.

	\subsection{Information Bottleneck in Decision-Level Fusion: Feature-Level Fusion vs. Logits Ensembling}
	
	To further assess whether cross-reservoir information is better exploited before or after class-level compression, we compare the proposed unified feature-level readout with an equal-weight decision-level logits fusion baseline. In the decision-level setting, each reservoir response block is first mapped by an independent local readout to a \(10\)-dimensional class-logit vector, and the final logits are obtained as
	\begin{equation}
		\label{eq:late_fusion}
		\bm z_{\rm LF}
		=
		\frac{\bm z_1+\bm z_2+\bm z_3}{3}.
	\end{equation}
	This late-fusion baseline achieves a full-test accuracy of \(96.84\%\), slightly above the strongest single-reservoir model but below the \(96.95\%\) accuracy obtained by the proposed feature-level fusion model.
	
	The observed difference is consistent with an early compression constraint, although the comparison does not establish that compression is the sole cause. In decision-level fusion, each \(768\)-dimensional reservoir block is first mapped by an independently trained readout to a \(10\)-dimensional logit vector. The final averaging operation can use only these compressed class-level representations. In feature-level fusion, a single classifier is instead optimized over all \(2304\) input features simultaneously.
	
	Importantly, the feature-level classifier remains linear and does not explicitly form cross-reservoir products or covariance features. Its advantage is therefore not that it directly evaluates quantities such as
	\(\phi_{K_r,i}\phi_{K_s,j}\), but that the class-feature weights of all reservoir blocks are optimized jointly under a common loss before reservoir-wise compression.
	
	To formalize the feature-level construction, the response blocks extracted from the three biased dynamical configurations are concatenated into
	\begin{equation}
		\label{eq:joint_representation}
		\begin{aligned}
			\bm{\Phi}(\bm I)
			&=
			\left[
			\phi_{\bm K_1}(\bm I),
			\phi_{\bm K_2}(\bm I),
			\phi_{\bm K_3}(\bm I)
			\right]
			\in \mathbb{R}^{9N},\\
			N&=256,\qquad 9N=2304 .
		\end{aligned}
	\end{equation}
	Before linear classification, the joint representation is standardized using training-set statistics:
	\begin{equation}
		\begin{aligned}
		\mu_j
		&=
		\frac{1}{S}
		\sum_{s=1}^{S}
		\Phi_j(\bm I_s), \\
		\widehat{\sigma}_j
		&=
		\left[
		\frac{1}{S-1}
		\sum_{s=1}^{S}
		\left(
		\Phi_j(\bm I_s)-\mu_j
		\right)^2
		\right]^{1/2}
		\end{aligned}
		\label{eq:training_feature_statistics}
	\end{equation}
	and
	\begin{equation}
		s_j
		=
		\begin{cases}
			\widehat{\sigma}_j,
			&
			\widehat{\sigma}_j\geq10^{-12},
			\\[3pt]
			1,
			&
			\widehat{\sigma}_j<10^{-12},
		\end{cases}
		\label{eq:feature_scale}
	\end{equation}
	and
	\begin{equation}
		\widetilde{\Phi}_j(\bm I)
		=
		\frac{\Phi_j(\bm I)-\mu_j}{s_j}.
		\label{eq:standardization}
	\end{equation}
	All feature means and standard deviations are computed from the training feature matrix only and are subsequently applied unchanged to the test features. The fallback \(s_j=1\) prevents division by a numerically constant feature channel.
	
	The classification logits are computed by a single linear readout,
	\begin{equation}
		\bm z(\bm I)
		=
		\bm W_{\mathrm{out}}
		\widetilde{\bm\Phi}(\bm I)
		+
		\bm g_{\mathrm{out}},
		\label{eq:linear_readout}
	\end{equation}
	where
	\(\bm W_{\mathrm{out}}\in\mathbb{R}^{C\times2304}\)
	is the readout matrix,
	\(\bm g_{\mathrm{out}}\in\mathbb{R}^{C}\)
	is the bias vector, and \(C=10\) is the number of classes.
	The predicted label is
	\begin{equation}
		\widehat y(\bm I)
		=
		\operatorname*{arg\,max}_{c\in\{1,\ldots,C\}}
		z_c(\bm I).
		\label{eq:prediction}
	\end{equation}
	
	Partitioning the readout matrix according to the three
	reservoir blocks,
	\begin{equation}
		\bm W_{\mathrm{out}}
		=
		\left[
		\bm W_1,\bm W_2,\bm W_3
		\right],
		\label{eq:partitioned_readout_matrix}
	\end{equation}
	gives
	\begin{equation}
		\bm z(\bm I)
		=
		\sum_{r=1}^{3}
		\bm W_r
		\widetilde{\phi}_{\bm K_r}(\bm I)
		+
		\bm g_{\mathrm{out}}.
		\label{eq:additive_feature_fusion}
	\end{equation}
	where
	\(\bm W_r\in\mathbb{R}^{C\times 3N}\)
	for \(r\in\{1,2,3\}\). Equation~\eqref{eq:additive_feature_fusion} makes clear that
	the classifier combines reservoir blocks additively. It jointly
	optimizes their class-feature coefficients, but does not explicitly
	introduce bilinear cross-reservoir interactions.
	
	The readout parameters are optimized by minimizing the empirical
	cross-entropy with \(L_2\) regularization:
	\begin{equation}
		\mathcal{L}
		=
		-
		\frac{1}{S}
		\sum_{s=1}^{S}
		\ln
		\left[
		\frac{
			\exp\!\left(z_{y_s}(\bm I_s)\right)
		}{
			\sum_{c=1}^{C}
			\exp\!\left(z_c(\bm I_s)\right)
		}
		\right]
		+
		\frac{\lambda}{2}
		\left\|
		\bm W_{\mathrm{out}}
		\right\|_F^2.
		\label{eq:loss_function}
	\end{equation}
	The bias vector \(\bm g_{\mathrm{out}}\) is not included in the
	\(L_2\) penalty. For all results in the main comparison,
	\(S=60000\), \(\lambda=10^{-3}\), the learning rate is
	\(0.05\), the batch size is \(128\), the number of epochs is
	\(150\), and the readout seed is \(20260621\).
	
	Because the readout is optimized on the uncompressed direct sum
	of the reservoir features, all class-feature coefficients are
	trained jointly under the same global loss. This construction
	relaxes the reservoir-wise compression constraint of independent
	local readouts, while remaining an additive linear classifier.

	\subsection{Relation to Generative Thermodynamic Computing}
	The present discriminative moment-resolved discriminative framework is complementary to recent generative thermodynamic computing. In generative thermodynamic computing, the trained Langevin system is required to transform noise into structured data, and the learning objective is naturally formulated in terms of reverse-trajectory likelihood and thermodynamic irreversibility. By contrast, the present work uses the Langevin system as a finite-time physical feature extractor. The input is already structured, and the computational question is how much discriminative information can be extracted from the transient nonequilibrium response distribution.
	
	This distinction leads to different computational objectives. Generative thermodynamic computing optimizes stochastic dynamics to transform noise into structured samples, whereas the present work uses finite-time trajectories to construct discriminative polynomial-moment features. The multi-reservoir results further indicate that reservoirs with distinct initialization and training histories can exhibit different classification-error overlaps. Thus, trajectory-level generative likelihood and moment-resolved finite-time readout represent complementary uses of Langevin dynamics for physical machine learning.
	
	\section{Conclusion}
	
	In summary, this work extends finite-time Langevin computing from mean-only readout to a moment-resolved representation based on the raw polynomial observables
	\(\mathbb{E}[\bm x]\),
	\(\mathbb{E}[\bm x^{\odot2}]\), and
	\(\mathbb{E}[\bm x^{\odot4}]\).
	These raw moments should not be interpreted as pure measurements of variance and kurtosis: they combine displacement and central-shape contributions while remaining naturally aligned with the polynomial structure of the onsite dynamics.
	
	Under the fixed MNIST reproduction protocol, the complete three-reservoir feature representation gives the best observed accuracy of \(96.95\%\). The difference from the strongest single-reservoir model is \(0.13\) percentage points and is not statistically significant under the reported exact McNemar test. The wrong-set Jaccard analysis shows that the reservoirs exhibit different degrees of classification-error overlap, but it does not by itself establish feature-space decorrelation or collinearity. The results therefore provide suggestive, rather than definitive, evidence that heterogeneous response bases can contribute complementary information.
	
	The current comparison also increases feature dimension, sampling resources, and readout capacity when additional reservoirs are introduced. Future work should include equal-resource controls, repeated trajectory and readout seeds, raw-versus-central moment comparisons, and observation-time or temperature scans. Such analyses will be needed to determine how much of the observed behavior originates specifically from dynamical heterogeneity, higher-order response statistics, and finite-time nonequilibrium evolution.

	\section{DATA AND CODE AVAILABILITY}
	
	The source code, fixed configuration files, random seeds, numerical protocols, and scripts used for the MNIST \(60000/10000\) reproduction are openly available at
	\url{https://github.com/djz0924/Nonlinear_Thermodynamic_Computer}.
	The accompanying release archive provides the \(K_1\), \(K_2\), and \(K_3\) reservoir weights, SHA256 checksums, final readout parameters, prediction files, software-environment metadata, and machine-readable result summaries. Large precomputed feature files are provided as optional release artifacts. The MNIST dataset is publicly available through its standard distribution channels and is not redistributed in the source repository.

	\bibliographystyle{apsrev4-2}
	\bibliography{references}
\end{document}